# $(l,s)-$Extension of Linear Codes

Axel Kohnert

*Lehrstuhl Mathematik II*
*Universität Bayreuth*
*95440 Bayreuth*
*Deutschland*

**Abstract**

We construct new linear codes with high minimum distance $d$. In at least $12$ cases these codes improve the minimum distance of the previously known best linear codes for fixed parameters $n, k$. Among these new codes there is an optimal ternary $[88, 8, 54]_3$ code.

We develop an algorithm, which starts with already good codes $C$, i.e. codes with high minimum distance $d$ for given length $n$ and dimension $k$ over the field $GF(q)$. The algorithm is based on the new defined $(l,s)-$extension. This is a generalization of the well-known method of adding a parity bit in the case of a binary linear code of odd minimum weight. $(l,s)-$extension tries to extend the generator matrix of $C$ by adding $l$ columns with the property that at least $s$ of the $l$ letters added to each of the codewords of minimum weight in $C$ are different from $0$. If one finds such columns the minimum distance of the extended code is $d + s$ provided that the second smallest weight in $C$ was $\geq d + s$. The question whether such columns exist can be settled using a Diophantine system of equations.

## Introduction

The most prominent example of an extension of a linear code which increases the minimum weight is the use of a parity bit in the case of a binary linear code of odd minimum weight. There is a series of papers where the authors try to generalize this situation.

A code is called *extendable*, if it is possible to find an extension which also increases the minimum distance. Extendability was studied by Hill and Lizak [11,12],

*Email address:* Axel.Kohnert@uni-bayreuth.de (Axel Kohnert).

van Eupen and Lisonek [19], Simonis [18] and in recent years by Maruta [14,15,16,17]. A common theme of this work is the study of the weight distribution of a linear code $C$. The authors derive certain conditions on the weight distribution which are sufficient for the extendability of the code.

We generalize this situation, as we no longer search for only one-step-extensions. We try to increase the length of the codewords by $l$ letters in a way such that the minimum distance increase by at least $1$. We call this a *good* extension.

This is different from previous work by Van Eupen and Lisonek [19] where they prove that in certain situations a ternary code is two-fold extendable, this says that it is possible to increase the length and also the minimum distance by $2$. The sufficient conditions ensure that the resulting code is self-orthogonal. Two-fold extendability was also studied in [15].

Concepts used but not defined in this text can be found in any book on linear codes (e.g. [1,2]).

## $(l, s)-$Extension

Let $C$ be a linear $[n, k]_q$ code of minimum distance $d$ with generator matrix $\Gamma$. We call this an $[n, k, d]_q$ code. $C$ is connected to its generator matrix $\Gamma$ via the relation:

$$C = \{v\Gamma : v \in GF(q)^k\}. \quad (1)$$

Let $c_1, \ldots, c_g$ be the codewords in $C$ of minimum weight $d$. There are vectors $v_1, \ldots, v_g$ from $GF(q)^k$ such that $c_i = v_i\Gamma$ for all the minimum weight codewords $c_i$. We call the set $V := \{v_1, \ldots, v_g\} \subset GF(q)^k$ the *minimum weight generator* of the code $C$. We are looking for an extension of the generator matrix $\Gamma$ by $l$ columns in a way such that the corresponding extended code has minimum distance $> d$. For an increase in the minimum distance it is necessary that all minimum weight codewords in $C$ are extended by at least one nonzero letter. This will be used to characterize a good extension.

The possible columns for the extension of the generator matrix are the non-zero vectors of $GF(q)^k$. We are interested in the minimum weight of the extended code, therefore we are only interested in the zero/non-zero property of the letters added to the codewords. This property is invariant under scalar multiplication of the possible column by a non-zero element from $GF(q)$, therefore we restrict to columns

$$\gamma_1, \ldots, \gamma_h \quad (2)$$

which are representatives of the one-dimensional subspaces of $GF(q)^k$. In order to have canonical representatives the first non-zero entry of $\gamma_i$ should be $1$. The number $h$ of possible canonical columns is $\frac{q^k-1}{q-1}$.

We have to check whether the extension by a possible column increases the weight of the actual minimum weight codewords. Again like in the case of the columns the minimum weight property is invariant under scalar multiplication by a non-zero element, therefore the number $s$ of the minimum weight codewords in $C$ is a multiple of $(q-1)$ and we have to check only $t := \frac{s}{q-1}$ elements from the minimum weight generator, which again are representatives

$$g_1, \dots, g_t \tag{3}$$

of certain one-dimensional subspaces of $GF(q)^k$. Here we also use canonical representatives.

For a systematic search by an algorithm define the *intersection matrix* $D$, which is a $t \times h$ matrix with entries equal to $0$ or $1$. The rows are labeled by the $t$ canonical representatives $g_1, \dots, g_t$ and the columns are labeled by the $h$ possible canonical columns $\gamma_1, \dots, \gamma_h$ . The entries are defined ($\langle, \rangle$denotes the inner product):

$$D_{i,j} := \begin{cases} 1 & \text{if } \langle g_i, \gamma_j \rangle \neq 0 \\ 0 & \text{if } \langle g_i, \gamma_j \rangle = 0 \end{cases} . \tag{4}$$

An entry $1$ at the position $i, j$ says that there is a non-zero letter in the codeword $c = g_i \Gamma'$ at position $m$ if a generator matrix $\Gamma'$ has $\gamma_j$ as the $m-$th column. An entry $0$ says that this letter is $0$. Using this we have the following theorem:

**Theorem 1** *good extension*

*Suppose $C$ is a linear $[n, k, d]_q$ code.*

*There is a code $C'$ with minimum distance at least $d+1$ built by $l-$fold extension of $C$, iff there are $l$ columns of the matrix $D$, such that for each row of $D$ there is at least one non-zero entry among the $l$ columns.*

**PROOF.** This equivalence is clear from the above description of the connection between the matrix $D$ and the encoding of the codewords via multiplication with a generator matrix.

We call such an $[n+l, k]_q$ code $C'$ with minimum distance $> d$ an $(l, 1)-$*extension* of $C$. We added $l$ columns to a generator matrix and got an increase of the minimum distance of at least $1$. The generator matrix of the code $C'$ is given by the extension of the generator matrix of $C$ by the columns corresponding to the selected $l$ columns of the matrix $D$. In the case of a gap of size $s$ between the minimum weight $d$ and the second smallest weight $d+s$ of $C$ we get:

**Corollary 2** $(l, s)-$*Extension*

*Let $C$ be a linear $[n, k, d]_q$ code with a second smallest weight $d+s$. We get an $[n+l, k]_q$ code $C'$ with minimum distance at least $d+s$ built by $l-$fold extension iff we can find a multiset of $l$ columns of the matrix $D$, such that for each row there are at a least $s$ non-zero entries among the $l$ columns.*

Such an extension is called an $(l, s)-$*extension*. In this corollary the multiset instead of a set (like in the above theorem) is necessary as it is possible that several copies of the same column are added to the generator matrix. The simplest case of an $(l, s)-$extension is the addition of a parity bit in the case of a binary code of odd minimum weight. This is an $(1, 1)-$extension.

For computational purpose we now state the problem as a Diophantine system of (in)equalities. We state this only for the case of an $(l, 1)$-extension, a more general version is also possible.

**Corollary 3** $(l, 1)-$*Extension as a Diophantine system of inequalities*

*Let $C$ be a linear $[n, k, d]_q$ code.*

*There is an $(l, 1)-$extension of $C$ iff there is a $0/1-$solution $x = (x_1, \ldots, x_h)^T$ of the following system of (in)equalities:*

$$\begin{array}{ll}(1)\ \sum x_i = & l \\ (2)\ \ Dx \ \geq & \begin{pmatrix} 1 \\ \vdots \\ 1 \end{pmatrix}.\end{array}$$

In a last step this system is changed into a Diophantine system of equations:

**Corollary 4** $(l, 1)-$*Extension as a Diophantine system of equations*

*Let $C$ be a linear $[n, k, d]_q$ code.*

*There is an $(l, 1)-$extension of $C$ iff there is a solution $\mathbf{x}|\mathbf{y} := (x_1, \ldots, x_h, y_1, \ldots, y_h)$ (with $x_i$ from $\{0, 1\}$, $y_i$ from $\{0, \ldots, l-1\}$) of the Diophantine system of equations:*

$$\left[\begin{array}{c|c} D & \begin{matrix} -1 & & & 0 \\ & \ddots & & \\ & & \ddots & \\ 0 & & & -1 \end{matrix} \\ \hline 1 \ldots 1 \ldots 1 \ldots 1 & 0 \ldots 0 \ldots 0 \ldots 0 \end{array}\right] \begin{pmatrix} \mathbf{x} \\ \mathbf{y} \end{pmatrix} = \left[\begin{array}{c} 1 \\ \vdots \\ \vdots \\ 1 \\ \hline l \end{array}\right]$$

From the values of the slack variables $y_1, \ldots, y_h$ it is possible to read off the effect of the new columns on the minimum codewords of the original code $C$. The weight of the $q - 1$ codewords corresponding to the row-label $g_i$ are increased by $y_i + 1$. Using this information together with the information on the second smallest weight it is possible to derive the number of minimum weight codewords in the code $C'$. The typical case is as follows: there is a slack variable $y_i = 0$, this means that the codewords in $C'$ corresponding to $g_i$ are of weight $d + 1$. It may happen (this is the even better case of an $(l, s + 1)-$extension), that all slack variables are at least $s > 0$, and the difference between the minimal weight in $C$ and the second smallest weight is $> s+1$, then the minimum distance of $C'$ is at least $d+s+1$, and again the number of codewords of minimum weight can be derived as above from the values of the slack variables. Above results are summarized in the following algorithm

**Algorithm 1** $(l,1)-$*extension*

***Input****: An $[n,k,d]_q$ code $C$ with generator matrix $\Gamma$, and a number $l > 0$.*

*Step 1: compute the canonical representatives $g_1,\ldots,g_t$ from the minimal weight generator.*

*Step 2: build the matrix $D$.*

*Step 3: try to find a solution $(x|y)$ of the Diophantine system of equations.*

***Output****: in the case of a solution $(x|y)$ there is an $[n+l,k]_q$ code with minimum distance $> d$.*

**Projective Geometry**

The connection between non-degenerate linear $[n,k]_q$ codes and finite projective geometry $PG(k-1,q)$ is well known (e.g. [2] p. 249). The matrix $D$ is a submatrix of the point-hyperplane incidence matrix of the projective geometry. We get the matrix $D$ from the original incidence matrix (points labeling the columns, hyperplanes labeling the rows) by taking only the hyperplanes orthogonal to the vectors from the canonical representatives of the minimum weight generator. Using this connection Maruta characterized the $1-$extendability of linear codes [14,17] using the intersection property between point and hyperplanes. This is equivalent to the study of the matrix $D$. For the more general case of $(l,1)-$extendability we get a generalization of these results and can restate the above theorem in terminology of projective geometry. Denote by $P$ the (multi-)set of points in $PG(k-1,q)$ corresponding to the non-degenerate $[n,k,d]_q$ code.

**Lemma 5** $(l,1)-$*extension in* $PG(k-1,q)$

*A $[n,k,d]_q$ code (with corresponding point-set $P$) can be extended to an $[n+l,k]_q$ with minimum distance $> d$, iff there is a set of $l$ points in $PG(k-1,q)$ such that the intersection number between $P$ and all the hyperplanes containing at least one of the $l$ points is $< n-d$.*

Working in finite projective geometries people restrict to projective codes, as in this case it is possible to work with sets of points instead of multisets. Algorithm 1 can be modified to generate only extensions which result to a projective code. For this remove the $n$ columns of $D$ corresponding to the points in the point-set $P$ of the code $C$. Now a solution of the corresponding Diophantine system of equations corresponds to a selection of points different from the points already in $P$.

**Comparison to other Methods, Limits of the Algorithm**

The $(l, s)-$extension is based on the well known extension of a generator matrix by one further column. But in general this does not increase the minimum distance of the code. One famous exception is the use of a parity bit, which is the simplest case of a $(1, 1)-$extension. The more general $(l, s)-$extension characterizes in which cases a generalization of this parity-bit method is possible.

The $(l, s)-$extension is an 'inverse' operation to the special puncturing described by Grassl and White in [10]. They remove $l$ columns of a generator matrix in a way such that each codeword of minimum weight has an entry equal to zero in at least $s$ columns of these $l$ columns. Given this property together with a condition on the second smallest weight the punctured code has minimum distance $d - l + s$. A code $C'$ constructed by $(l, s)-$extension of a code $C$ will give back $C'$ using special puncturing. This is the generalization of the well known pair (extension - puncturing) for linear codes.

The typical case in which one may try to apply the $(l, 1)-$extension is a sequence of codes of lengths $n, n + 1, \dots, n + l$ where the best known minimum distance is constant say $d$. This is a hint that the extension by a single column does not work for the codes of length $n, \dots, n - l + 1$, but $(l, 1)-$extension applied to the code of length $n$ may work. All the new examples in the final section were found starting with such a situation.

The Diophantine system of equations can be solved quite effectively by Wassermann's implementation of the LLL-algorithm [21]. Another method in the case of small $l$ is the complete enumeration of all $l-$tuples of column indices. As this problem is a covering problem one may hope to use Knuth's dancing links program [13]. This only solves exact cover problems, we also tried a modified version which uses the fast original data structure of the dancing links program.

The size of the computational problem is given by the size of the matrix $D$. A linear code $C$, where we can apply algorithm 1 has about $5000$ codewords. An exceptional situation is a code where the size of the minimum weight generator is small, i.e. the coefficient $A_d$ of the weight enumerator $w_C$ is small. The algorithm allows to handle problems in the same size like other proposed algorithms (e.g. Q-extension [4], special puncturing [10], descent method [3]) used for the search of codes with improved minimum distance $d$ for fixed parameters $n, k, q$.

To apply $(l, s)-$extension to a code $C$, we need to know the minimum weight generator of the code. It is known [20] that already the computation of the minimum weight (which is less information) is $NP-$hard. The same is true if we want to use other ([11,14,16,17,19]) extendability results, where information about the weight enumerator is necessary.

If we try $(l, s)-$extension on a code, constructed using the methods described in [5,6,7,8], we already got during this construction the representatives of the minimum weight generator. On the other hand codes which can be handled using this method are in most cases small enough, and to compute the minimum weight gen-erator using complete enumeration or more sophisticated algorithms based on advanced methods for the computation of the minimum distance [1,10] is 'cheap' compared to the time necessary to run algorithm 1.

An obvious generalization is to take into account not only the words of minimum weight $d$ but also words with weight $d+1,\ldots,d+s$. Building the corresponding intersection matrix together with the system of equations would allow to look for an $(l,s)-$extension $(s>1)$ in the case where there is no gap of size $s$ between the two smallest weights.

## Example and Results

We found a *new* $[82,8,49]_{q=3}$ code, which is a $(2,1)-$extension of a previously computed $[80,8,48]_3$ code with $1320$ codewords of minimum weight. The corresponding Diophantine system of equations has $(3^8-1)/2=3280$ variables (=possible columns for extension) and $1320/2 = 760$ equations. Among all possible pairs of columns we found a covering pair of possible generator matrix columns.

This new code can be extended twice using $(1,1)-$extension, giving new $[83,8,50]_3$ and $[84,8,51]_3$ codes. For the last one it was again possible to apply $(2,1)-$extension followed by an $(1,1)-$extension giving new $[86,8,52]_3$ and $[87,8,53]_3$ codes and a last $(1,1)-$extension finally ended at an *optimal* (no larger minimum distance is possible for these values of $n,k,q$) $[88,8,54]_3$ code. This is a self-orthogonal code, but the last twofold extension is not covered by the theorem [19] of van Eupen and Lisonek, as there are codewords of weight 2 mod 3 in the $[86,8,52]_3$ code.

Other recently found codes using $(l,s)-$extension have the following parameters:

$[132,8,81]_3$, $[197,6,142]_4$, $[212,6,153]_4$, $[227,6,165]_4$, $[232,6,169]_4$, $[242,6,177]_4$, $[247,6,181]_4$.

All these codes are improvements of Brouwers on-line table [9] of codes, where one can look up the largest known minimum distance for given triples of $(n,k,q)$ together with the best known upper bound for the maximum possible minimum distance.